# From Acquaintances to Friends: Homophily and Learning in Networks

BY MIHAELA VAN DER SCHAAR AND SIMPSON ZHANG[1]

This paper considers the evolution of a network in a discrete time, stochastic setting in which agents learn about each other through repeated interactions and maintain/break links on the basis of what they learn from these interactions. Agents have homophilous preferences and limited capacity, so they maintain links with others who are learned to be similar to themselves and cut links to others who are learned to be dissimilar to themselves. Thus learning influences the evolution of the network, but learning is imperfect so the evolution is stochastic. Homophily matters. Higher levels of homophily decrease the (average) number of links that agents form. However, the effect of homophily is anomalous: mutually beneficial links may be dropped before learning is completed, thereby resulting in sparser networks and less clustering than under complete information. There may be big differences between the networks that emerge under complete and incomplete information. Homophily matters here as well: initially, greater levels of homophily increase the difference between the complete and incomplete information networks, but sufficiently high levels of homophily eventually decrease the difference. Complete and incomplete information networks differ the most when the degree of homophily is intermediate. With multiple stages of life, the effects of incomplete information are large initially but fade somewhat over time.

## 1. Introduction

This paper constructs a novel model of network formation. Our model is the first that makes it possible to analyze the joint impact of homophily (the desire of agents to link with other agents that are similar to themselves), incomplete information and learning on network formation, and the first that allows quantitative predictions about the probabilities that given networks will emerge and not (as in all previous literature) simply qualitative predictions about which (very special) networks <u>might</u> emerge.

Homophily in agent preferences is one of the most important determinants of a network's structure. Agents have limited time and resources to spend with their connections, so they will often prioritize links with those that are the most similar to themselves. As the well-known sociology paper by McPherson et al (2001) explains, "personal networks are homogeneous with regard to many sociodemographic, behavioral, and intrapersonal characteristics." Furthermore, "network studies showed substantial homophily by demographic characteristics such as age, sex, race/ethnicity, and education". In addition, not only do agents tend to form more links with others that are similar to themselves, but "ties between nonsimilar individuals also dissolve at a higher rate." Thus homophily has a strong influence on all stages of network development, from the formation, to the maintenance, to the eventual dissolution of links in networks.

Homophily is closely related to the concept of social distance, which describes the perceived distance between different groups of society. Such differences could be based on social class, levels of income or choices of occupation. As the strength of homophily increases, a society becomes more stratified since agents prefer to link more closely with those that are similar to themselves. As Marsden (1988) states, "Strong homophily tendencies imply less frequent intergroup relations and thus impede integration." Since in many situations, "the class structure is neither desirable nor inevitable" (Laumann and Senter 1976), it can be an important objective to reduce the extent of stratification. This work analyzes how incomplete information and learning may reduce the level of stratification in certain situations. It has also been recognized in the sociology literature that although homophily may cause global integration to decrease, local integration may increase since agents interact more strongly with others in their own group. As Marsden (1988) explains, "homophily tendencies…are at once indicators of decreased global and increased local integration". In our model homophily results in more local integration when information is complete and agents know more of each other, but less global integration due to each agent's capacity constraints causing them to sever links with less similar agents.

---

[1] Mihaela van der Schaar: mihaela@ee.ucla.edu
Simpson Zhang: simpsonzhang@ucla.edu
The authors would like to thank Ahmed Alaa for his valuable assistance with this project.

Although agents can vary drastically in their characteristics, and such heterogeneity is of great importance for the network structure, the exact types of each agent are often unknown initially. This is especially true when agents are meeting and forming networks for the first time. Each agent must therefore learn about other agents over time through mutual interactions. Since any learning process is noisy and imperfect, the agents may not learn each other's types with complete accuracy, so mutually beneficial links may be dropped by mistake, leading to sparser networks between the agents. For instance, two agents who were in actuality very similar in type may sever their link due to errors in the learning process. Such an outcome has two effects: first each agent now has one less link in the network, but second each agent may try to replace this link by seeking out other agents who are not as close to themselves. Thus the diversity of links in the final network may be improved, and the network may exhibit less clustering of links.[2]

The main aim of this paper is to capture this interaction between learning, network formation, and homophily. We consider a model wherein agents start with incomplete information about other agents, learn about each other through mutual interactions to resolve this uncertainty, and exhibit homophily by choosing to form connections only with agents who are believed to be similar and to maintain connections only with agents that they learn are actually similar. Our results show that incomplete information can lead to lower clustering than in networks that would arise under complete information. The reason is that due to errors in the learning process, agents may not actually find out who their most similar neighbors truly are. Thus they may actually end up deciding to link with agents that are not actually the closest in type, and so more diverse connections would be formed within the network, but with lower clustering overall.

Consider a group of individuals interacting with each other in an online setting, such as through e-mail, instant messaging, or a website. Agents interact with each other by, for instance, sending messages, pictures, or web links. Due to time constraints, each agent can maintain only a fixed number of links, and due to homophily, these agents would prefer to remain connected with the other agents that are most similar to them in type. As time progresses, individuals *learn* more information about other individuals, *update* their beliefs about the type of each agent, and *change* their linking decisions as a result. Thus learning by the agents causes the network topology to evolve over time as more information is revealed. The end result of the learning process could be a network that is more heterogeneous than in offline settings where learning is more accurate. Agents may link with other agents that are less similar to themselves, and the resulting online networks may be more diverse due to the incomplete information, resulting in less clustering in the final network. Some support for this can be seen in Kossinets and Watts (2006), which uses data from e-mail interactions in a university setting to infer social ties among students and concludes that "homophily with respect to individual attributes appears to play a weaker role than might be expected".

---

[2] Clustering is a measure of how often the neighbors of any given node are also connected with each other. Although in some network papers clustering is a result of triadic closure, it has also been shown to arise in network models featuring homophily, such as Gauer and Landwehr (2015), because the friend of your friend is likely to be close in type to you as well. Our results show that incomplete information tends to undercut this effect.

We may think of the above process of forming links as making acquaintances and the process of maintaining links as turning acquaintances into friends. At the end of this process (which might be thought of as a "stage of life" – kindergarten, public school etc.), agent characteristics may change and the process of forming and maintaining links may repeat. The new links may or may not include some of the old links: the friends of childhood may or may not be the friends of adulthood. Agents who have lost previous friends may actively expand their initial circle of acquaintances in hopes of finding more friends. We show via simulations that the effect of incomplete information is large in the initial phases of life, but eventually decrease in later phases as the network becomes more similar to the complete information network.

In summary, our paper represents a novel contribution to the network formation literature in several dimensions. It is the first to analyze the joint impact of homophily and incomplete information in a network formation setting. Moreover, it is the first to offer a model which makes quantitative predictions about the probabilities that given networks <u>will</u> emerge and not just qualitative predictions about which very special networks (e.g. stars, core-periphery) <u>might</u> emerge.

## 2. Literature Review

Previous papers in the network literature studying network formation usually considered settings of complete information where agents perfectly know each other's types. For example, the papers by Jackson and Wolinsky (1996), Bala and Goyal (2000), Watts (2001), and Galeotti and Goyal (2010) all consider networks where the agents have complete information. In these models, since agents are aware of the exact types of all other agents there is no learning. The network formation dynamics arise instead from externalities and indirect benefits between agents that are not directly linked, such as in the connections model of Jackson and Wolinsky (1996). For some networks, such as informational networks, these indirect benefits are important, as an agent who has many neighbors will likely produce higher quality information as well. However, in other networks such as friendship networks where interactions are more self-contained, these indirect benefits are less relevant and it is the type of each direct neighbor that matters the most. The network formation process that arises in such situations therefore centers more on incomplete information and agent learning than on changes in the value of indirect benefits.

We do not assume any indirect benefits in our model and focus instead on the effect of incomplete information and learning by the agents. Agent learning has a strong influence on the network formation process. Due to the noisiness in the learning process, agents may not learn their neighbors' types perfectly. As a result, their final linking decisions may not be with those agents who are in actuality the closest to them in type, but instead with other agents who are further away in type. This has implications for the final shape of the network because agents may exhibit less clustering in their final links.

Existing network papers that have analyzed the effects of homophily on network structure have thus far not considered the impact of incomplete information. One such paper is Johnson and Gilles (2000), which adds homophily to the Jackson and Wolinsky (1996) connections model. They model each agent as occupying a location in a social space, with each agent preferring to link with other agents that are at close social distances. They show that such homophily has a large impact on the network structure, causing the network to be more densely connected and clustered than otherwise. The network structure in our model is also largely driven by homophily, but incomplete information and learning play a crucial role and may undercut the effects of homophily in certain settings.

The recent paper by Iijima and Kamada (2014) also considers the impact of social distance on network structure. As in our paper, agents prefer to link with closer agents and use a cutoff rule for linking decisions. However, their paper measures social distance using a special "k'th norm" metric, which considers the shortest $k$ distances among $m$ total dimensions. They show that when agents care about more dimensions network clustering as well as the average path length increase. In contrast, we use a Euclidean metric to measure the social distance, and agents have incomplete information about each other and learn about each other. Moreover, agents have capacity constraints which inform their linking decisions.

Two recent papers also consider network formation in settings in which agents have continuous types. Gauer and Landwehr (2015) assume that formation follows a Bernoulli Random Graph model in which the probability of a link between two agents is greater when agent types are closer together. There is no learning and no active decision making by agents; linking is entirely a random process. Baccara and Yariv (2015) consider the strategic formation of groups for the joint production of public goods. In their model, agents have different interests and face free riding issues after the group forms, which impacts the resulting group structures. Again, information is complete and there is no learning.

Theoretical models (by e.g. Currarini et al (2009) and Bramoulle et al (2012)) about homophily have also been proposed which consider discrete types. Since these papers consider finite and discrete types, there is no natural metric space over each type and homophily is measured instead by the number of links between the same and different type agents. Our approach thus has the advantage of allowing for a natural metric space over the types, allowing us to analyze the distance of connections within a network. These papers also consider complete information and do not analyze the impact of learning about agent types in the network formation process.

Our paper is also related to the recent literature on learning in networks. Recent work by Golub and Jackson (2010), Golub and Jackson (2012) and Acemoglu et al (2011) analyzes observational learning in social networks. However, all these papers suppose that there is a fixed exogenous network on which the agents interact, and that agents learn about an exogenous state of the world via this network by observing their neighbor's actions. The papers show results on the speed and accuracy of the learning that can be achieved by agents connected using different networks. Our paper is significantly different because agents learn about other agent types instead of an exogenous state of the world. As such, agents will wish to update their linking decisions over time as their beliefs about the agents with whom they are connected change. Thus, in our paper the network and learning co-evolve and the emerging networks evolve endogenously unlike these works where the network evolves exogenously.

The paper most closely related to ours is the recent paper by Song and van der Schaar (2015). Like us, this paper also considers learning by agents about the types of other agents in the network, and it shows how incomplete information and the learning process can lead to a wide variety of network structures and dynamics. However, this paper considers preferences that are homogeneous across agents, instead of the current paper in which agents exhibit homophily-based preferences and connect with other agents that are close in type to them. In addition, while this paper shows that the set of network structures and number of connections tend to increase with incomplete information, our results show that incomplete information tends to have a negative effect on the number of links. Furthermore, we show results on clustering and network diameter.

## 3. Model

There are a finite set of $I$ agents who can form undirected links with each other over a network. Agents are heterogeneous and vary according to their types, which for instance could be interpreted as their age, race, personality, hobbies, interests, etc. Each agent $i$ has a privately known type $\theta_i$, and this type is drawn from a commonly known normal distribution $N(\mu_i^0, \sigma_i^2)$.[3] This distribution represents the specific population group that agent $i$ is coming from, which incorporates all the publically known information about the agent such as the agent's location or gender. This distribution can vary across agents if they come from different populations, or it may be the same if they all come from one population.[4] Thus, $\mu_i^0$ and $\sigma_i^2$ may be agent-specific or may be common across agents. The mean of the distribution represents the central belief about the agent and the variance of the distribution represents how informative the initial public belief is.

The network is formed by the agents based on their beliefs about other agents' types. We assume that the links are formed via bilateral consent and can be severed unilaterally. Due to homophily, agents will link only with other agents that have prior types that are close to their own true type. Specifically, we assume that agent $i$ will want to be linked with agent $j$ if and only if $|\theta_i - \mu_{ij}| < \delta_i$, where $\mu_{ij}$ is the mean of agent $i$'s current belief about agent $j$'s type and $\delta_i$ represents the tolerance threshold of agent $i$. Such a linking rule could arise if each agent's utility was strictly decreasing in the distance between it and any other agent's type, as in Iijima and Kamada (2014). For instance, suppose that each agent $i$'s utility for a link with agent $j$ were given by $-|\theta_i - \mu_{ij}| + \delta_i$ and the utility of not linking were 0. Then if each agent acted myopically, it would only link with other agents whose expected types were less than a distance of $\delta_i$ away. We note that $\delta_i$ can be interpreted in terms of an agent's preferences, or more broadly as a social norm that prevents agents from linking with others that are too far away. Based on this linking rule, each agent makes its initial linking decision (on the basis of its initial beliefs $\mu_{ij}^0$), and from these linking decisions the network $G^0$ is formed, which we call the *baseline network*. The baseline network describes the connectivity at time $t = 0$.[5]

After forming the baseline network $G^0$, agents will continue to interact with each other, and through this process they will learn about each other's types, update their beliefs and change their linking decisions. Interactions in our model take place in two phases: the *experimentation phase* and the *link refinement phase*. These two phases can occur repeatedly many times; repeated occurrences could be interpreted as part of the life cycle – kindergarten, public school, college, etc. We first describe the model with one iteration of these phases, and we then discuss the implications of multiple stages of life afterwards.

In the experimentation phase, agents learn about the neighbors that they are linked with. This experimentation phase consists of $T$ distinct *action periods* in which agents receive signals about their neighbors, update their beliefs in a Bayesian fashion, and revise their linking decisions. At the end of each period $t$, each agent $i$ receives a signal about each agent $j$ that it is linked with, and these signals $s_{ij}^t$ are normally distributed with mean $\theta_j$ (the true type of agent $j$) and variance $\sigma_{ij}^2$. We assume that signals are conditionally independent across agents and links.[6] We denote the history of signals agent $i$ receives about agent $j$ up to time $t$ as $\mathcal{H}_{ij}^t =$ $\left\{s_{ij}^{t'}\right\}_{t'=1}^{t'=t}$. Based on these signals, agent $i$ decides whether or not to remain linked with agent $j$, and the link survives until the next period if and only if both agents agree to remain linked.

---

[3] We note that many of our results are robust to the specific distributional assumptions. The normal distribution is used here for greater tractability in the Bayes updating process. However, other distributions can also be used. For illustration purposes, we analyze a bimodal distribution in Section 9 and show that our results can be extended to this setting.

[4] Whether one or multiple population groups is appropriate depends on the detail of the initially known public information. For instance, in settings where very little is known about the agents, such as in an online website where no public details are available, the agents can be assumed to be coming from a single population group. However if the website publically lists the nationality of each agent, then there would be multiple population groups, with a group for each nation.

[5] The baseline network $G^0$ can be thought of as being formed during a meeting process by the agents, in a classroom, event, or gathering for instance.

[6] This assumption entails that agents are interacting across each link bilaterally, so that the information that is revealed from each interaction is private instead of public.

The linking decision in each action period is made on the basis of Bayes rule, as an agent $i$ Bayes updates the prior belief of any agent $j$ that it is connected with. Agent $i$'s posterior belief about agent $j$'s type at the end of period $t$ given its observations of agent $j$ will be normally distributed with mean $\mu_{ij}^t = E[\theta_j|\mathcal{H}_{ij}^t]$ and variance $\sigma_{ij}^2(t)$. As above, we assume that due to homophily agents prefer to link with other agents that have a type that is similar to their own. Specifically, agent $i$ will remain linked with agent $j$ if and only if $\left|\theta_i - \mu_{ij}^t\right| < \delta_i$, where $\delta_i$ is the tolerance threshold of agent $i$. If however $\mu_{ij}^t$ is too far away from $\theta_i$, then agent $i$ will cut off the link with agent $j$. After this link is cut off, no more signals will be received in future periods and so learning will cease. Given the decisions of all agents, at the end of period $t$ the network $G^t$ will be formed.

After these $T$ action periods, the agents will no longer update their links until they learn all the remaining information about their neighbors. For instance, this could result from the agents becoming more patient after the initial uncertainty has decreased somewhat and thus no longer updating their links based on intermediate signals. We model this by assuming that each agent $i$ receives a final signal (or multiple composite signals) about each neighbor $j$ that it is still linked with, and that this signal is perfectly informative and resolves all remaining uncertainty. After finding out the other agent's type, agent $i$ will decide to remain linked with agent $j$ if only if the true type of agent $j$ is within the threshold $\delta_i$, that is $\left|\theta_i - \theta_j\right| < \delta_i$. Agent $j$ will act according to the same rule, and so agents $i$ and $j$ will be linked at the end of the experimentation phase if and only if $\left|\theta_i - \theta_j\right| < \min(\delta_i, \delta_j)$. Based on the agents' decisions, the *learned information network* $G^L$ will be formed. Notice that this is not necessarily the network that would be formed if every agent knew the true type of every other agent: agents who are not linked in $G^0$ because of initial beliefs, or who learn incorrectly about each other from the signals and so cut off links, will never receive the final signal that reveals each other's true types and hence never link even if they would want to link given their true types.

Learning in the experimentation phase could also be modeled in continuous time. In such a model agents would learn about each other via a continuous time diffusion process: $dB_{ij}(t) = \theta_i dt + \sigma_{ij}^2 dZ_{ij}(t)$, where $Z_{ij}(t)$ is a standard Brownian motion (independent over all pairs of agents $i$ and $j$) and $B_{ij}(t)$ represents all the information that agent $i$ has revealed to agent $j$ up to time $t$. Although learning is continuous, it is natural to assume that linking decisions take place at discrete times, and that the agents act based upon the information received between decision times. A property of Brownian motion is that it is normally distributed when sampled at discrete time intervals. If we assume that agents update their links a total of $T$ times during the experimentation phase, and these updates occur at time intervals $1, 2, 3, \ldots T$, then the information that the agents receive would be identical to the information received in the discrete time model. Thus the Bayes updating performed by the agents and the distribution over beliefs of the agents will also be identical. And after $T$, if the agents were to wait an arbitrarily long length of time before updating their links again, then the agents would end up with arbitrarily precise information, similar to the discrete time model with the perfectly revealing final signal. Thus the continuous time and discrete time models would result in equivalent information as well as equivalent linking decisions.

Although the continuous time model would yield exactly the same conclusions as our discrete time model, the continuous time interpretation makes clearer the impact of adding additional stages of actions (increasing $T$) in the discrete time model. Because all information is eventually revealed perfectly anyway, additional stages of actions in the experimentation phase do not increase the total amount of information revealed, but instead gives agents more opportunities to sever their links. Therefore these additional opportunities only serve to undercut the amount of learning. Additional times that actions are allowable would decrease the amount of overall learning in the model instead of increasing it.

In the second phase, the *link refinement phase*, agents will form the final network based on their capacity constraints. In many real life settings, agents have limited time and so there is a maximum on the number of links they can maintain. We model this by imposing a capacity constraint on each agent $c_i \in \mathbb{N}$. We assume that agents first go through the learning stage described above, and they then apply the capacity constraint only after all the information has been revealed at the end of the first phase. This is a reasonable assumption as agents are often willing to link with more neighbors while they are still experimenting, but once the information has been learned they will narrow their choices based on other considerations. While in the experimentation phase links are made on a bilateral basis between different agents, the capacity phase serves to couple together all the links across the network.

The capacity constraint is implemented in the following manner. After the experimentation phase is over, each agent will know the true type of its remaining neighbors. If every agent has fewer neighbors than their respective capacities, the learned network $G^L$ will also be the *final network* $G^*$. Otherwise some agents will break off links with other agents in order to meet their capacity requirements. We do not explicitly model the process by which this happens, but instead we will require that the final network that is formed be pairwise stable. Pairwise stability in this model implies that if agents $i$ and $j$ are linked in the network $G^L$ but not in the final network $G^*$, then agents $i$ and $j$ would not prefer to form a link with each other, by breaking other links in $G^*$ if needed to satisfy their capacity constraints[7]. We show below that there is always a unique pairwise stable network in our model.

Figure 1 shows the complete sequence of events for our model.

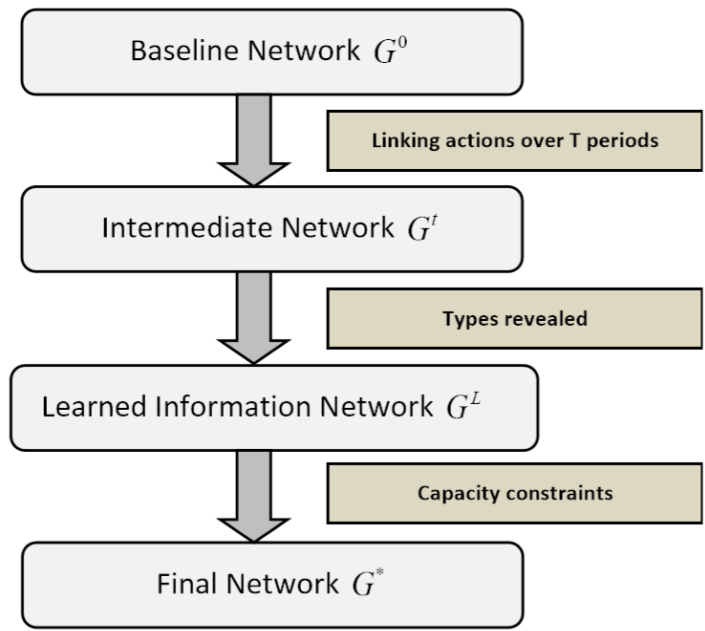

Figure 1: Sequence of Events in Model

The two phases in our model can also be iterated. This could represent agents changing their types due to both internal and external factors as they pass through different stages of their lives. Formally, each agent's true type can be subjected to random normal noise after the final network $G^*$ is formed, so that the type of agent $i$ becomes $\theta_i' = \theta_i + \varepsilon$, $\varepsilon \sim N(a, b)$. Thus each agent's true type will once again have a normal distribution from the perspective of the other agents. The phases can be iterated using the new associated priors. Note that the priors may no longer be common knowledge among agents, as they will be depend on what was learned in the previous stage of network formation. Agents who were linked to agent $i$ in $G^L$ and know agent $i$'s old type perfectly will now have a belief regarding agent $i$'s new type $\theta_i' \sim N(\theta_i + a, b)$. On the other hand, an agent $j$ who did not learn agent $i$'s type fully and instead has a posterior belief of $N(\mu_{ij}^t, \sigma_{ij}^2)$ will now have a belief $\theta_i' \sim N(\mu_{ij} + a, \sigma_{ij}^2 + b)$. However the analysis below will still remain similar. In this way a long term evolution of the network can also be studied. We will present results for the iterated stages of life in a later section.

## 4. Belief Updating and Network Dynamics

In this section we analyze in more detail the learning that occurs and the networks that can result from the experimentation periods. Since each agent's type has a normal prior distribution and the signals are also normal, the agent's posterior type distribution after each period will also be normally distributed. Given the assumptions above, we can use standard methods to find the expectation of agent $i$'s first period belief of agent $j$'s type, $\mu_{ij}^1$, as a function of the signals $s_{ij}^1$ and agent $j$'s initial mean $\mu_j^0$ and variance $\sigma_j^2$: $\mu_{ij}^1 = \frac{\mu_j^0/\sigma_j^2 + s_{ij}^1/\sigma_{ij}^2}{1/\sigma_i^2 + 1/\sigma_{ij}^2}$.

The belief at the end of period $t$, $\mu_{ij}^t$, can be found by iterating this formula with the new signals, as long as agents $i$ and $j$ are still connected. However, if the expected type at the end of period $t$ is greater than a distance of $\delta_i$ beyond $\theta_i$, then agent $i$ will cut off the link and the agents would no longer be connected. Then learning will not occur in the next period and the true type of agent $j$ would not be revealed to agent $i$ no matter what the true type is. Similarly if agent $j$'s belief about agent $i$ becomes a distance of $\delta_j$ away from agent $j$'s true type at the end of period $t$, the link would also be cut off and a similar situation would result.

The probability that agent $i$'s belief about agent $j$'s type is still within a range of $\delta_i$ at the end of period $t$ conditional on being connected in period $t-1$ can be calculated explicitly using standard mathematical properties of normal distributions. The following technical lemma provides an explicit expression for the survival probability fixing the previous period mean of agent $i$'s belief about agent $j$, $\mu_{ij}^{t-1}$, denoted $P(S_{ij}^1 | \theta_i, \theta_j, \delta_i, \mu_{ij}^{t-1})$. The ex ante probability that a link still exists in period $t$ can then be found by using this formula iteratively. This probability is shown to be strictly increasing in the agent tolerance $\delta_i$. In the following lemma $\Phi$ represents the standard normal cdf function.

LEMMA 1:

$$P\left(S_{ij}^t \middle| \theta_i, \theta_j, \delta_i, \mu_{ij}^{t-1}\right)$$

$$= \Phi\left(-\frac{\tau_0\left(\mu_j^{t-1} - \theta_i - \delta_i\right)}{\sqrt{\tau_r}} + \sqrt{\tau_r}(\theta_i + \delta_i - \theta_j)\right)$$

$$- \Phi\left(-\frac{\tau_0\left(\mu_j^{t-1} - \theta_i + \delta_i\right)}{\sqrt{\tau_r}} + \sqrt{\tau_r}(\theta_i - \delta_i - \theta_j)\right)$$

$$\text{where } \tau_0 = \frac{1}{\sigma_j^2(t)}, \tau_r = \frac{1}{\sigma_{ij}^2(t)}$$

PROOF: The probability $P\left(S_{ij}^t \middle| \theta_i, \theta_j, \delta_i, \mu_i^{t-1}, \mu_j^{t-1}\right)$ is equal to $P\left(s_{ij} \geq \left(-\frac{\tau_0\left(\mu_j^t - \theta_i + \delta_i\right)}{\tau_r} + (\theta_i - \delta_i)\right), s_{ij} \leq \left(-\frac{\tau_0\left(\mu_j^t - \theta_i - \delta_i\right)}{\tau_r} + (\theta_i + \delta_i)\right)\right)$. Normalize the outputs by subtracting the mean and dividing by the standard deviation to get the probability in terms of the standard normal signal $Z$: $P\left(Z \geq -\frac{\tau_0\left(\mu_j^{t-1} - \theta_i + \delta_i\right)}{\sqrt{\tau_r}} + \sqrt{\tau_r}(\theta_i - \delta_i - \theta_j), \ Z \leq -\frac{\tau_0\left(\mu_j^{t-1} - \theta_i - \delta_i\right)}{\sqrt{\tau_r}} + \sqrt{\tau_r}(\theta_i + \delta_i - \theta_j)\right).$ ∎

Using the above result, we can get an explicit formula for the probability that any specific network emerges at the end of each period. (Note that this is the first time in the theoretical network literature in which such probabilities are derived for an incomplete information setting.) We can fix a realization of types for the network and then use Lemma 1 to find the probability that two agents $i$ and $j$ are linked at the end of the each period if they are linked in the baseline network $G^0$. Since we assumed that all signals are independent, the probability that agents $i$ and $j$ remain linked at the end of the first period is the product of the probability that neither wishes to break off the link, or $P\left(S_{ij}^1 \middle| \theta_i, \theta_j, \delta_i\right) * P\left(S_{ji}^1 \middle| \theta_j, \theta_i, \delta_j\right)$. Then, since signals are independent across pairs of agents, we can do a similar calculation for every pair of agents linked in the baseline network to see if they are linked at the end of period 1. Finally, we can integrate over all of the possible type realizations for the network, $\{\theta_i\}_{i \in I}$, to get the *ex ante* probability that any network emerges at the end of period 1. Notice that for each subsequent learning period up until period $T$ we can repeat the above procedure by taking conditional probabilities to get the probability that any network $G^t$ is formed. For instance, the probability that agents $i$ and $j$ are still linked at the end of period 2 is given by:

$$\iint_{\mu_i, \mu_j} P\left(S_{ij}^t \middle| \theta_j, \theta_j, \delta_i, \mu_{ij}^1\right) * P\left(S_{ji}^t \middle| \theta_j, \theta_i, \delta_j, \mu_{ji}^1\right) dF\left(\mu_i^1 \middle| \theta_i\right) dF\left(\mu_j^1 \middle| \theta_j\right)$$

In the above formula $F\left(\mu_i^1 \middle| \theta_i\right)$ represents the cdf of the mean of the belief about agent $i$ at the end of the first period.

Using the above methodology, we can next analyze what types of networks can emerge at the end of the experimentation phase, after the final perfectly revealing signal has been sent and all of the information has been learned. To understand which networks $G^L$ can emerge, we investigate whether a link $l_{ij}$ between agents $i, j$ can exist at the end of the experimentation phase. This depends on the baseline network $G^0$. If two agents $i$ and $j$ are not initial neighbors (i.e. $g_{ij}^0 = 0$), then it is certain that $g_{ij}^L = 0$. If two agents $i$ and $j$ are initial neighbors (i.e. $g_{ij}^0 = 1$), then the existence of this link $l_{ij}$ requires that the expected types of both $i$ and $j$ do not move too far in the wrong direction, so that neither agent wishes to break off the link. Hence $G^L$ will always be a subset of the baseline network $G^0$, and is composed only of links between agents whose expected types do not move too far away from the other agent's true type.

To calculate the exact probability that two agents are linked at the end of the first phase, we can utilize the linking probability given above in Lemma 1. We know that Lemma 1 gives the probability that agents are still linked after each period, and the probability that they will still be linked after the first phase conditional on being linked after period $T$ is equal to 1 if $|\theta_i - \theta_j| < \min(\delta_i, \delta_j)$, and is equal to 0 otherwise. Thus the ex ante probability that agents $i$ and $j$ are still linked at the end of the first phase will be strictly smaller than at the end of each period $t$, and can be found by integrating the probability of being connected at the end of period $T$ over all pairs of types that are close enough together.

## 5. Effect of Capacity Constraints

Now we move on to the analysis of the second phase, the *link refinement phase*. As stated, in the second phase each agent has full information regarding their remaining neighbors and will make linking decisions according to their own capacity constraints. We analyze the set of final pairwise stable networks that can emerge. Pairwise stability means that no agent can deviate and form a link with another agent that it is linked with at the start of the round, such that both agents are now linked with closer agents than before while still satisfying their respective capacities. It turns out that there is almost always a unique pairwise stable network in our model (with probability measure zero there may be ties between agent types, resulting in non-uniqueness).

LEMMA 2:

There almost always exists a unique pairwise stable network.

PROOF:

Note that with probability 1, the distance between the true types of any two agents will be different across each pair of agents. Suppose that this property is satisfied, and that there exist two different pairwise stable networks $G$ and $G'$. Let agents $i$ and $j$ be the closest in type pair of agents such that a link exists between them, without loss of generality, in $G$ but not in $G'$. Let the distance between these two agents be $d$. Then since $G$ is a pairwise stable network, either agent $i$ or agent $j$ must be at capacity in $G'$ and linked with agents that are all a distance strictly less than $d$ away. But then not all of these links could exist in network $G$, since agent $i$ and $j$ are linked in $G$ and thus each agent can form a maximum of $c_i - 1$ links with other agents. Thus either agent $i$ or $j$ must be linked with another agent less than a distance of $d$ away in network $G'$ but not in network $G$. This contradicts the hypothesis that the link between agents $i$ and $j$ is the minimum distance link that is different in networks $G$ and $G'$.

∎

We abstract away from the exact process in which the pairwise stable network can be reached. However we note that the pairwise stable network can be found by using the following procedure.

*Stable Network Computation Procedure:*

First, consider all links in the complete information network $G^L$, and select the link with the smallest distance between the two agents' types. If these two agents are under capacity, then link these two agents in the final network $G^*$. Then, remove this link from the network $G^L$ and select the new link with the smallest distance between the two agent types. Repeat this process, and if any agent has its capacity filled, remove that agent and all the agent's links from $G^L$ when selecting future agents. This procedure must terminate since the number of rounds is limited by the total number of links in the stable network. In addition, the resulting matching will be pairwise stable, since all the agents that are linked in the final network have the shortest possible distances among all links in the learned information network, subject to the capacity constraints.

We can show some properties of the stable network procedure given above, which we proved maps a learned information network $G^L$ into a unique final network $G^*$. These results concern the impact of changes in the tolerance levels of each agent on the final network that is formed. The theorem assumes that a network $G^L$ is already given, and so does not consider the impact of $\delta_i$ on $G^L$ itself.[8]

THEOREM 1:

The following properties hold for the stable network procedure that leads to a final network given a learned information network $G^L$:

1. Increasing $\delta_i$ has no effect on $i$'s links if $i$ is at capacity already
2. Increasing $\delta_i$ cannot affect $i$'s links to agents closer than $\delta_i$
3. Increasing $\delta_i$ below capacity weakly increases the $i$'s links and the average distance
4. If all agents have the same $\delta_i$, raising $\delta_i$ for a single agent has no impact.
5. Raising $\delta_i$ for an agent with neighbors that have tolerance less than $\delta_i$ cannot change the final network.

PROOF:

1. Fixing the network $G^L$, increasing $\delta_i$ does not increase any of the links that agent $i$ would wish to form if it is already at capacity. Suppose to the contrary that agent $i$ now forms a link with agent $j$. Then the distance between agent $i$ and agent $j$ must be less than the maximum distance between agent $i$ and any of its previous links since agent $i$ was at capacity. But then agent $i$ and agent $j$ would already have been linked under the previous capacity, a contradiction.
2. We assume that the agent is not at capacity, because otherwise the statement follows from 1. If an agent gets rid of a link with a distance less than its original tolerance, it must now be at capacity and have all links that are less than the distance of the link it removed. But then the agent would have been at capacity originally as well, a contradiction.
3. From 2 we know that increasing the tolerance cannot affect any of agent $i$'s links to agents that are at a distance less than the original tolerance. Thus agent $i$ will not change its previous links and may form more links if it was not originally at capacity, but these links must be at a higher distance.
4. By 2, we know that none of the agent's links with agents less than a distance of $\delta_i$ can change. But since all agents have the same tolerance, agent $i$ cannot form any links of distance higher than $\delta_i$ either. Therefore the result follows.

---

[8] The joint impact of $\delta_i$ on $G^L$ and $G^*$ will be discussed in the next section.

5. This follows from a similar reasoning as 4.

∎

Note that these results were proved fixing a learned information network $G^L$. However, because $\delta_i$ affects experimentation, a higher $\delta_i$ leads to more experimentation and more links in $G^L$. Therefore these results may not be true ex ante, before the experimentation phase, due to the impact on $G^L$ itself. However, when agents have complete information about each other these results will always be true. With incomplete information, the effect on the agent's number of links and linking distance in the final network may be ambiguous as we show below. We will also include numerical simulations showing in which cases an agent's average linking distance increases and which cases it decreases for specific networks.

## 6. Final Network Properties

In this section, we synthesize the previous discussions and state properties of the final networks $G^*$ that will emerge. A few properties are immediate. First of all, any link that does not get started, that is either $|\theta_i - \mu_j^0| < \delta_i$ or $|\theta_j - \mu_i^0| < \delta_j$ is not satisfied, cannot be part of the final network since no link could ever form between such pairs of agents. Secondly, all agents have their types learned perfectly at the end of the first phase. Thus for any pairs of agents such that $|\theta_i - \theta_j| < \min\{\delta_i, \delta_j\}$, no link can exist between them in the final network. So for agents that are linked in the final network $G^*$ there are restrictions on both the initial expected types of agents, and on the true types of agents. For pairs of agents that satisfy both of these requirements, a link in the final network is still not guaranteed, as the expected type of any agent can still move too far away from the true type in the experimentation phase due to noise. Finally, the capacity constraint that is applied in the link refinement phase requires that agents $i$ and $j$ have large enough capacities and not too many links with even closer agents.

The final capacity effect serves to limit the diversity of the network's links, and with complete information agents would end up prioritizing links with those that they are closest to and ignoring agents that are less similar to themselves. Under incomplete information however, there is a positive probability that agents may form links to neighbors that are at a greater social distance away from themselves than under complete information. Furthermore, the greater the initial variance of agent types, the greater the probability that the true types of agents are further away from each other as well. Thus both effects allow for links between agents that are further away, and we show below that the level of clustering in the network tends to decrease.

### Impact on Final Network of Removing Links from $G^L$

We now analyze the relationship between $G^L$ and $G^*$, specifically the effect of a removal of links from a given learned information network $G^L$, and we show the potential complications that can arise. To analyze the effect of a removal of links, we need to analyze the stable network procedure presented above in more detail. As mentioned, there is a mapping between the learned information network and the final network. This is because there is (almost always) a unique pairwise stable network. We now analyze this mapping in more detail, specifically the effect of a removal of links between agents from the learned information network on the resulting final network.

We consider the simplest case where each agent has capacity of one link and so can form a link with only one other agent. We show that even in this case complications can arise. In this simple case, each agent $i$ will want to link with the agent $j$ closest to itself in the network $G^L$. If that agent $j$ wishes to link to a different agent, then the first agent $i$ will want to link with the agent $k$ that is the second closest, and so forth. We consider an arbitrary network $G^L$, and we analyze the effect of removing a single link $g_{ij}^L$ from this network.

Suppose without loss of generality that $\theta_i < \theta_j$ and that the social distance of this link $g_{ij}^L$ (the distance between agent types) is $d$. First notice that unless this link is in the final network, such a removal has no impact on the final network. Now suppose that this link is in the final network. All links of social distance less than $d$ in the final network cannot be affected by this removal based on the description of the procedure above. Links of social distance $d$ or above may be affected. The first impact is on agents $i$ and $j$. Both agents will now be connected with an agent further away from themselves in the final network (if they are still connected). Suppose that agent $i$ is now connected with an agent $k$ in the final network, and that agent $j$ is now connected with an agent $l$ in the final network. Then either $k$ or $l$ must now have a shorter link than before. In fact both must have a shorter link by pairwise stability unless $\theta_k$ and $\theta_m$ are both larger than $\theta_j$ or smaller than $\theta_i$. In that case one of these agents may have a longer link. Thus while agents $i$ and $j$ are hurt by the removal of their link, their new partners will be hurt in some cases and will benefit in some cases. The effect on the average distance of all links in the network is ambiguous. It could be the case that the average distance increases or decreases depending on the types of other agents. This is summarized in the following theorem.

THEOREM 2:

Suppose all agents have a capacity of $1$. If a link between agents $i$ and $j$ is removed from $G^L$, then the distance of the links of $i$ and $j$ must increase if they are still linked. The average distance of all links in the network could either decrease or increase.

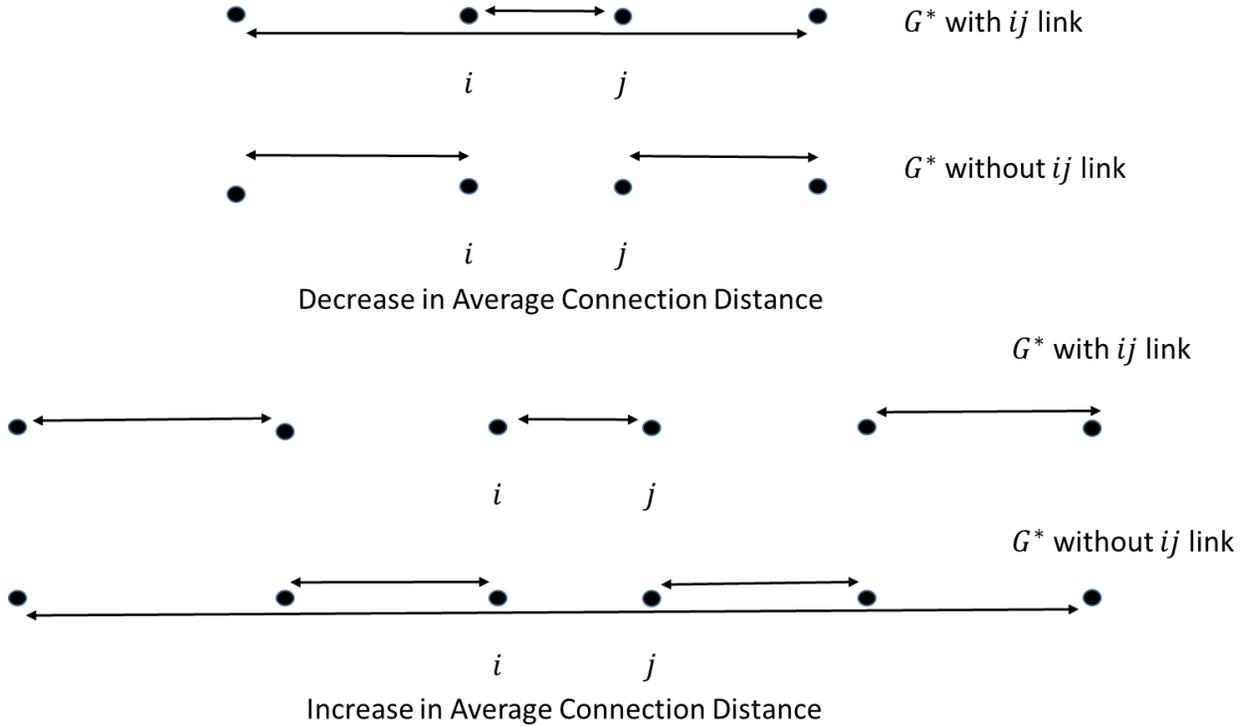

Figure 2: Ambiguous Effect of Link Removal on Average Connection Distance

Figure 2 shows both an example where the average distance of links in the final network increases and one in which it decreases. Consider two different $G^L$, one which is a clique and one which is a clique with the link between agents $i$ and $j$ removed. In the top figure with four agents, the average connection distance in the network increases from the first to second $G^L$ network, whereas in the bottom figure the average connection distance decreases. This indeterminacy of what happens to links in the final network when links in the intermediate networks are removed means that the effect of extra information on average connection distance is indeterminate as well. This creates difficulties in getting tractable analytical expressions for the probabilities that different networks emerge, so we will proceed in the subsequent sections by employing simulations to analyze the forces involved.

## 7. Homogenous Population Distribution Example

We now impose a few restrictions on the model parameters to focus on a specific setting. We assume that all of the $I$ agents have types drawn from the same initial prior distribution $N(\mu^0, \sigma^2)$. Such an assumption is satisfied when the agents all have similar backgrounds that are not distinguishable - for instance each agent is an anonymous member of some online forum. With this common prior assumption, we can immediately derive some results about the network formation process in each time period. We also assume for simplicity that $T = 1$, so there is one period of learning. With more stages of learning, as discussed previously there would be more links cut off in the incomplete information network and thus an even larger difference from complete information for the network $G^L$. We analyze the specific dynamics in each period, starting with what happens before the first period.

*Before First Period*

Notice that in this setting only agents with a true type that is within $\delta_i$ of the population mean $\mu^0$ will form any initial links. Agents that fail this criteria know that their true type is too far away from the average expected type, and so will not want to experiment with any other agent. We call such agents *self-isolated* agents. Such agents will never form any links and remain as singletons in all subsequent time periods. The failure of these self-isolated agents to form links may be undesirable from a social standpoint, as there is a positive probability that another agent was within $\delta_i$ of a self-isolated agent, and a link could have been sustained between that pair. But due to the initial beliefs of these self-isolated agents, they do not wish to experiment and find other agents like themselves.

Agents that do have a true type within $\delta_i$ of $\mu^0$ will form links with all other such agents. We call these agents *participating* agents, because they are close enough to the population mean that they expect a high chance of linking with a partner similar to themselves. At the start of the first period, the network $G^0$ will be a clique composed of participating agents, with self-isolated agents as singletons.

A special case that we can consider is when all agents have the same homophily preferences and use the same linking cutoff $\delta$. In this case we will also be able to analyze comparative statics of the model with respect to the size of this cutoff. Note that the $\delta$ in this case can be interpreted as a social norm, since a larger $\delta$ means that all agents are more tolerant of those that are different from themselves while a smaller $\delta$ means that the agents are less tolerant. Our first result, Lemma 3, shows that more tolerant social norms will lead to more participating agents and less isolation from the network. The lemma follows from the discussion above. Thus to prevent isolation of agents, a more tolerant social norm should be adopted initially.

LEMMA 3:

The larger the size of $\delta$, the smaller the set of self-isolated agents and the larger the set of participating agents in the network.

*First Period Dynamics*

During the first period, signals will be sent and the agents will update their beliefs based on these signals. For each neighbor, an agent will stay linked only if the expected belief about that agent's $\mu_j^1$ is within $\delta_i$ of their own type. The probability of this occurring is given in Lemma 1 as:

$$P\left(S_{ij}^1 \middle| \theta_i, \theta_j, \delta_i\right) = \Phi\left(-\frac{\tau_0(\mu^0 - \theta_i - \delta_i)}{\sqrt{\tau_r}} + \sqrt{\tau_r}(\theta_i + \delta_i - \theta_j)\right)$$
$$- \Phi\left(-\frac{\tau_0(\mu^0 - \theta_i + \delta_i)}{\sqrt{\tau_r}} + \sqrt{\tau_r}(\theta_i - \delta_i - \theta_j)\right)$$

Note that based on the initial linking actions that form $G^0$, we know that $\theta_i$ and $\theta_j$ are separated by at most $\delta_i + \delta_j$ due to the triangle inequality. We do not assume that the agents take this into account when Bayes updating; instead, we assume that the agents Bayes update merely use the signal that they receive. However, when calculating the *ex ante* probability that any network emerges at the end of period 1, this fact should be taken into account. The *ex ante* probability that any link between agents $i$ and $j$ exists at the end of period 1 will be given by:

$$\int_{\mu^0 - \delta_j}^{\mu^0 + \delta_j} \int_{\mu^0 - \delta_i}^{\mu^0 + \delta_i} P\left(S_{ij}^1 \middle| \theta_i, \theta_j, \delta_i\right) * P\left(S_{ji}^1 \middle| \theta_j, \theta_i, \delta_j\right) d\Phi(\theta_i) d\Phi(\theta_j)$$

*After First Period Dynamics*

After the first period, agents that are linked in $G^1$ will learn all remaining information about each other and find out each other's true type. The probability that a link remains is $1$ if the agents are within $\delta_{ij} \equiv \min(\delta_i, \delta_j)$. Therefore, we can get the *ex ante* probability that a link remains between agents $i$ and $j$ in the complete information network as:

$$\int_{\mu^0 - \delta_j}^{\mu^0 + \delta_j} \int_{\max(\theta_j - \delta_{ij}, \mu^0 - \delta_i)}^{\min(\theta_i + \delta_{ij}, \mu^0 + \delta_i)} P(S_{ij}^1 | \theta_i, \theta_j, \delta_i) * P(S_{ji}^1 | \theta_j, \theta_i, \delta_j) \, d\Phi(\theta_i) d\Phi(\theta_j)$$

This probability is strictly smaller than the first period probability, and the network learned information $G^L$ will be a subnetwork of $G^1$.

*Link Refinement Phase Dynamics*

In the second phase, agents that are still linked in $G^L$ will evaluate their current links on the basis of capacity constraints. Agents will cut off links if they are over capacity, and a final stable network will develop as discussed above. Fixing the vector of true types, we can map any network $G^L$ into a final network $G^*$ using the stable network procedure that was described previously. The *ex ante* probability of a final network $G^*$ occurring can be found by using the above formula to get the probability that a network $G^L$ and a corresponding true type vector that leads to $G^*$ occurs, and adding up the probabilities of all these networks to get the probability of $G^*$.

We wish to make comparisons between the final network $G^*$ that develops from the learning process against a network that would develop if information were complete. Notice that the effect of the learning process is to make some links unviable. That is, there are some pairs of agents whose true types were closer than $\delta_{ij}$ and may have wanted to link in the complete information network, but due to errors in the learning process ended up not being linked in $G^L$. In this sense, the learned information network $G^L$ will always be a subnetwork of the complete information network $G^C$, which is defined as the network between all pairs of agents with types closer than $\delta_{ij}$ ignoring capacity constraints.[9] Thus we want to study the effect of a removal of links from $G^C$ on the final network $G^*$, where the removal is due to errors in learning in the experimentation phase. We will do so via Monte Carlo simulations. We note that numerical integration techniques can also be used to get these results, since we have the analytical expressions above.

## 8. Numerical Simulations and Asymptotic Theorems

In this section we will present several figures showing the results of numerical simulations for the homogeneous population network. Refer to the technical appendix for exact details on how the simulations are run as well as specific definitions of all the metrics that are used. We assume that there are 10 agents that all come from a homogeneous population, and we show the impact of a change in social tolerance on various metrics of the network. Figure 3 shows the Jaccard distance between the final network with complete information, which we call $G_C^*$, and final network with incomplete information $G^*$ as a function of the social tolerance. The Jaccard distance measures the difference in the links of the complete information and the incomplete information network. There are two different forces that affect the Jaccard distance as social tolerance increases. First a larger social tolerance tends to cause more possible links, allowing for more places in which the complete and incomplete information networks differ. This effect serves to increase the Jaccard distance. On the other hand, a larger social tolerance means that learning is more accurate, and this effect serves to decrease the Jaccard distance. Interestingly, the simulations show that there is a maximum value of this distance at an intermediate level of social tolerance, and the distance goes to zero as the social tolerance becomes very large or very small.

---

[9] This is the network that would be formed if all agents knew all other agents' types perfectly initially and there were no capacity constraints.

We can prove that the learned information networks and the complete information network will with high probability be very similar if the tolerance is either very small or very large. If the tolerance is very small, no links form in either case. If the tolerance is very large, all links form initially and few are broken. The following result formalizes this intuition.

THEOREM 3:

As $\delta \to \infty$ or $\delta \to 0$, the probability that the learned information network $G^L$ is the same as the complete information network $G^C$ converges to $1$ and the Jaccard distance between the final network with complete information $G_C^*$ and final network with incomplete information $G^*$ converges to 0 in probability.

PROOF:

In the limit as the tolerance goes to zero, we will have a completely disconnected network for any ex post realization of types under both complete and incomplete information, so the result holds. In the limit as the tolerance goes to infinity, the amount of learning becomes perfect, and so links in $G^L$ will almost surely be the same as the links in $G^C$. The Jaccard distance also converges to zero in probability as a direct result of the fact that in both limits the links in $G^L$ and $G^C$ will be the same almost surely. Thus the capacity constraints will be applied in the same way for both networks, resulting in the same final networks.

∎

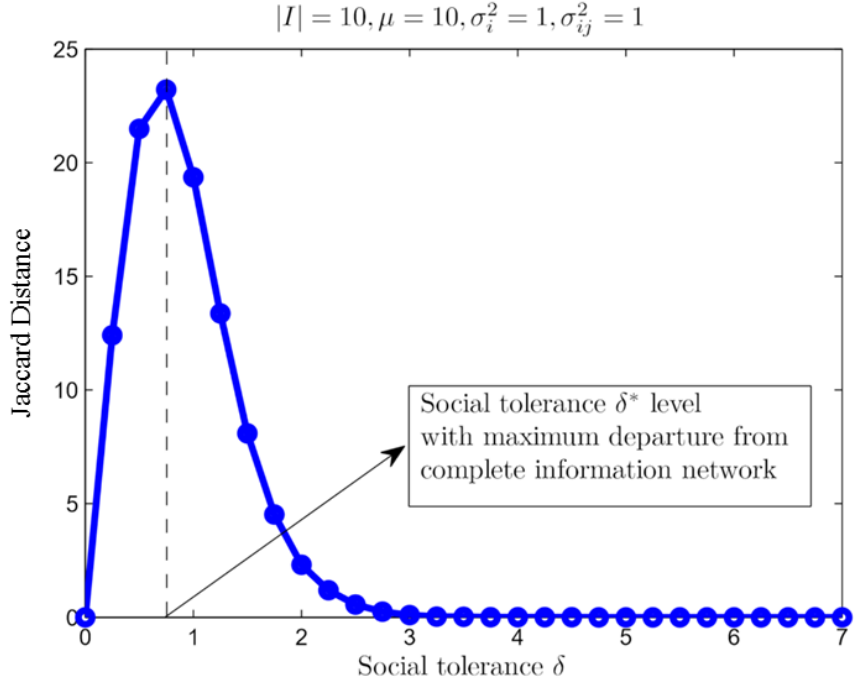

Figure 3: Jaccard Distance between the complete and incomplete information network as a function of the social tolerance

We next analyze how the average number of links per agent in the final network $G^*$ depends on the social tolerance. It might seem intuitive that as social tolerance increases, the number of links per agent should also increase. However, this may not be true for every realization of signals and types. To see why, consider two social tolerance levels, $\delta' < \delta$. It is true that the set of links in $G^L$ under $\delta'$ must be a superset of the set of links under $\delta$ for every realization. (This is because if a link would have been cut off under $\delta'$, it would have been cut off under $\delta$ as well.) However, the capacity constraints, which govern which links in $G^L$ survive in the passage to the final network $G^*$, may interfere with additional links being passed to the final network.

This is highlighted in the example in Figure 4, which the potential effects of adding links to $G^L$. Suppose that all agents shown in these figures have a capacity of one. We note that agents $i$ and $j$ have the shortest social distance among all agents, and so would form a link in the final network if they have a link in $G^L$. In addition, a link is not possible between agents $l$ and $m$ in either the top or bottom $G^*$ network since such a link is not present in either $G^L$ network. In the top $G^L$ network a link is not present between agents $i$ and $j$, and so agent $i$ links with agent $l$, while agent $j$ links with agent $m$ in that $G^*$. In the bottom $G^L$ network this link is present, and so agent $i$ links with $j$, but agents $l$ and $m$ do not link. Then in the top $G^*$ final network there will be two links formed, but in the bottom $G^*$ final network only a single link will be formed. Thus the number of links formed in the final network reduces given an increase in the number of links in $G^L$.

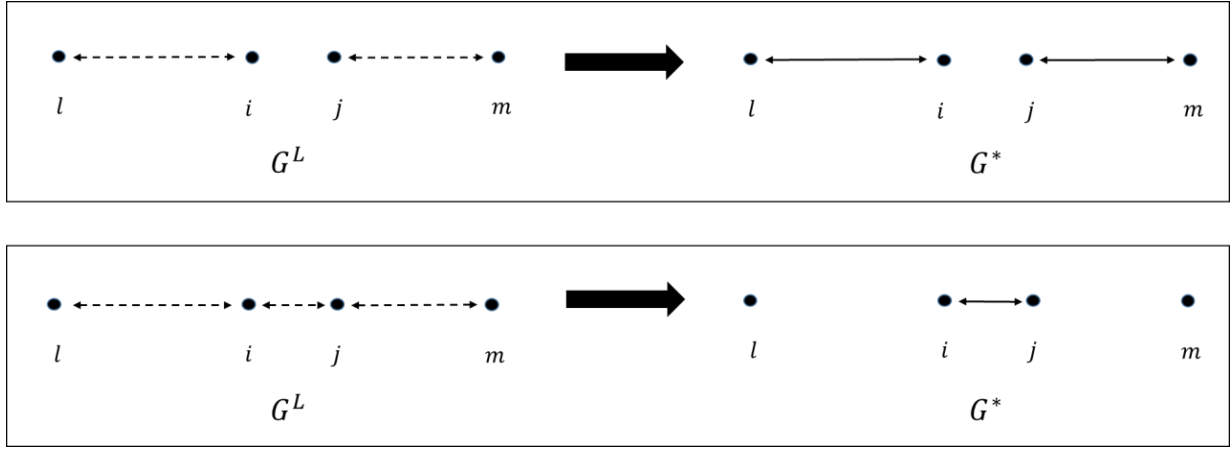

Figure 4: Decrease in total number of links in $G^*$ as a link between the agents $i$ and $j$ is added to $G^L$

Since it is possible for an increase in links in $G^L$ to decrease the links in $G^*$, the ex post effect of a higher social tolerance $\delta$ is ambiguous. Figure 5 suggests however that the average effect is to increase the total links in the network. Thus the increase in the links of $G^L$ dominates any potential issues that the implementation of the capacity constraints could cause.

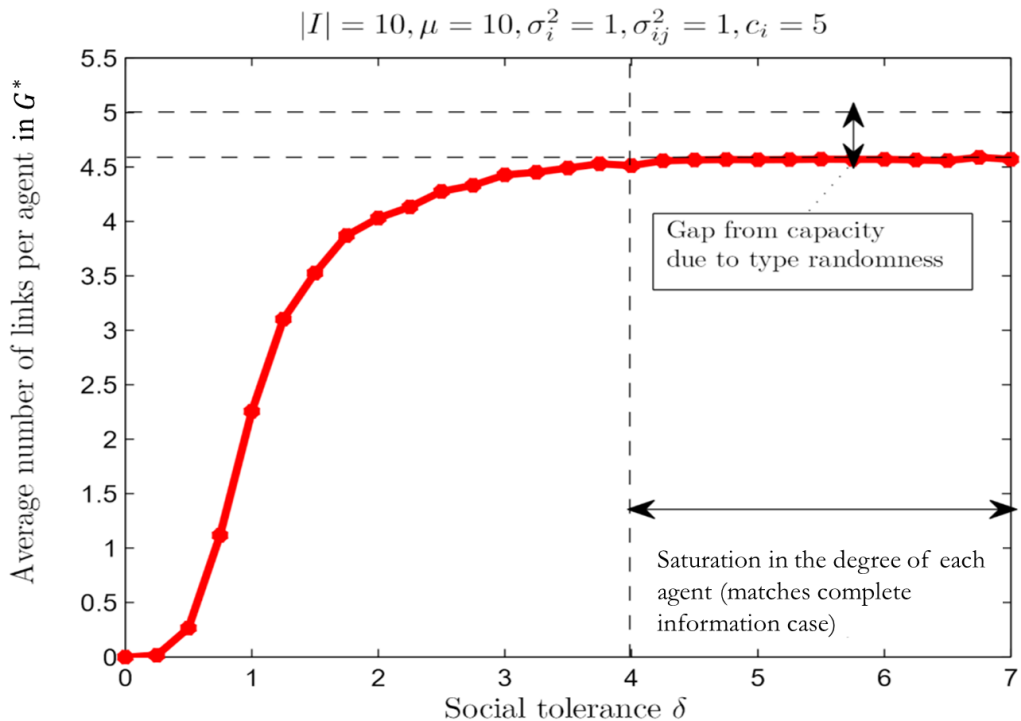

Figure 5: Average Number of Links Per Agent in Final Network Based on Social Tolerance

Figure 6 shows the average total distance of links per agent as a function of the social tolerance. This is defined as the average of the sum of the distance in types between each agent and all its neighbors. It can be observed that as the social tolerance increases, the number of links increases as well. However, the normalized average distance, equal to the sum distance divided by the social tolerance, increases to a maximum and then decreases. The average distance is zero at low $\delta$ since links become very unlikely, and the normalized distance is also zero which shows that the rate at which the average distance goes to zero is faster than the rate at which $\delta$ goes to zero. Thus it should be increasing in the beginning. However, at high $\delta$ the sum distance should saturate since greater social tolerance does not have an impact once the capacity level is hit.

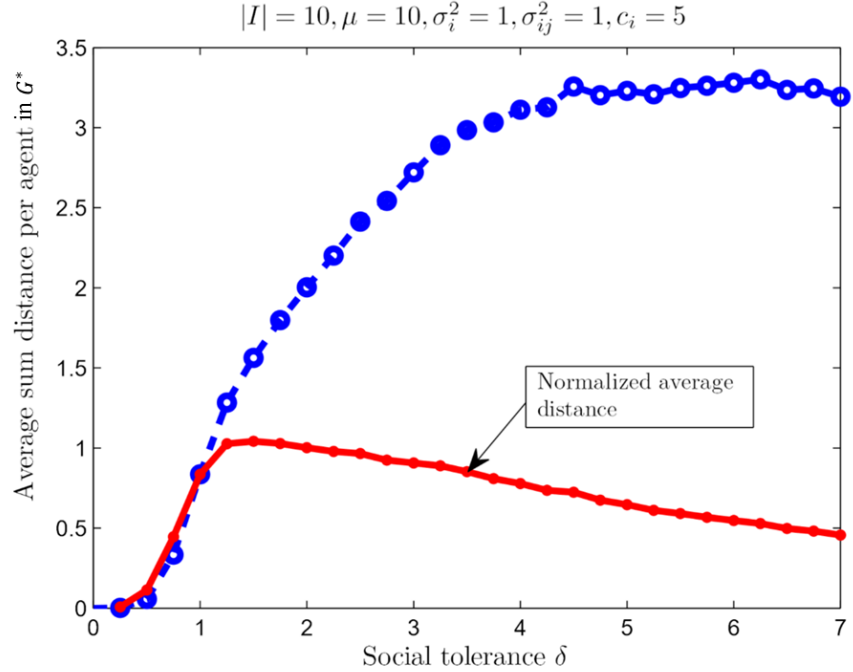

Figure 6: Average Distance in Incomplete Information Final Network as a Function of Social Tolerance

We now show the effect of social tolerance on clustering within the network. Clustering is a measure of how frequently a neighbor of a node is connected with other neighbors of the same node, or how often friends of friends are connected. We analyze the average of the local clustering coefficient of all the nodes of the network. The local clustering coefficient of a node is equal to the proportion of links that exist among the subgraph of that node and its direct neighbors divided by the total number of links that are possible within this subgraph.

Figure 7 shows the difference in the average local clustering coefficient between the final complete information network $G_C^*$ and incomplete information network $G^*$. The complete information network always exhibits greater clustering than the incomplete information network. This is due to the fact that links are more likely to be dropped in the learning process under incomplete information. However, when the social tolerance is very high or low, the difference goes to zero, which is in line with the above simulation showing that the Jaccard distance also goes to zero as the social tolerance gets very large or small. In fact this is a direct corollary of Theorem 3 above. Overall the results show that the effect of incomplete information is to lower the overall level of clustering in the network, which counteracts the increase in clustering that homophily tends to bring.

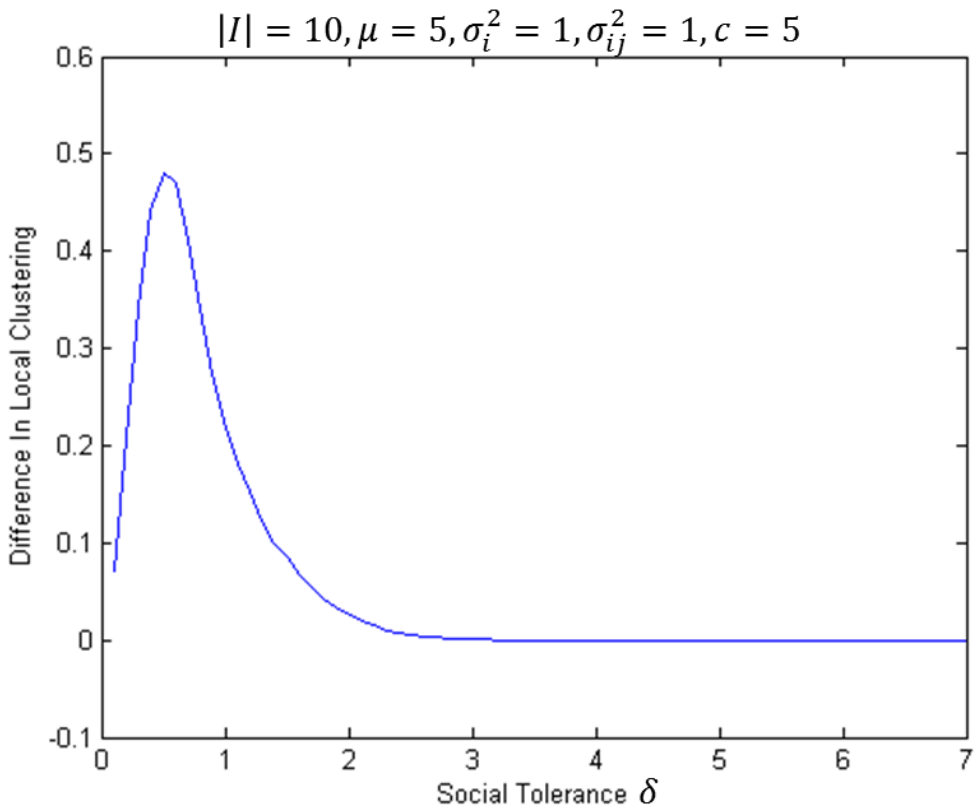

Figure 7: Difference in Local Clustering as a Function of Social Tolerance

## 9. Extensions

*Extreme Agent Networks*

An interesting setting to analyze is when most of the agents have types that are very close together, but a few agents exhibit extreme types. For instance, we might have a single central group (e.g. a moderate political party) with fringe groups (e.g. extreme factions) on the sides. The network would represent the links that are formed by such political groups, and our results would show when extreme factions can form links with the center versus being isolated. We can formalize this by assuming that there are $K$ central agents, whose types lie within the interval $[\alpha, \beta]$, where the length of the interval is equal to $d$, and $I - K$ extreme agents, whose types lie outside of this interval and a distance of at least $d$ away from both endpoints $\alpha$ and $\beta$, i.e. within the interval $(-\infty, \alpha - d] \cup [\beta + d, \infty)$. This ensures that all the agents of non-extreme types would prefer to link with any other agents of a non-extreme type than an extreme agent. Figure 8 shows an example of such a network with two extreme agents.

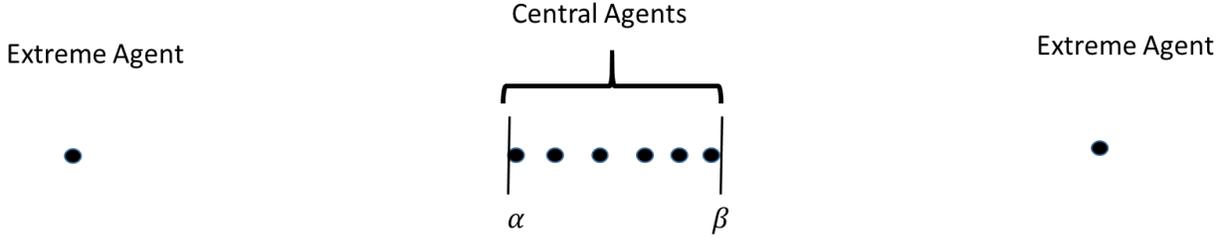

Figure 8: Example of Extreme Networks

When can an extreme agent connect with a central agent? First it is necessary for all central agents to have large enough social tolerances that they are willing to connect with the extreme agents. Even if this condition holds, there must still be a central agent that is under capacity after considering linking with all other central agents, since we have assumed that the extreme agents are all farther away from the central agents than any central agent is from another central agent. If there is a central agent that is under capacity after considering all other central agents, we say that there exists a central agent *with an available link*. We can prove a result about when such an event is possible and when it will not be possible.

THEOREM 4:

Suppose $\delta > d$. If the capacity is $1$ for the central agents then there will be $1$ available link if the number of central agents, $K$, is odd and no available links if it is even. If the capacity is $K - 1$ then there will be no available links. For all other capacity values, there is always a positive probability that a final network emerges in which there are available links.

PROOF:

For the first result when the capacity is equal to $1$, notice that by the assumption that the types of the extreme agents are all farther away than the maximum possible type difference of the central agents, all the central agents will link with each other first before linking with an extreme agent. If there is an even number of central agents then the central agents will all form pairs with each other, resulting in no available links. If there are an odd number of extreme agents, then there will be one central agent without a partner, and thus 1 available link.

If the capacity is equal to $K - 1$, then the only possible network for the central agents is a clique in which all the central agents are linked with each other. This is because no matter what the true types of the central agents are, given the capacity value they will be able to link with every other central agent. After doing so, there will be no available links left over for the extreme agents.

Finally suppose the capacity is something other than the above two values. If the capacity is greater than or equal to $K$, there will always be available links as even after forming a clique with other central agents each central agent will still not be at capacity. Suppose then that the capacity is strictly less than $K - 1$ and strictly larger than $1$. We show that there exist networks that emerge with positive probability in which the central agents have available links.

First assume that the number of central agents $K$ is not divisible by $c + 1$. In such a case, let $m$ be the greatest integer such that $m(c + 1) < K$, and let $r = K - m(c + 1)$. We give a distribution of real types for the central agents such that the resulting final network will have an available link if $G^L$ is a clique amongst the central agents, which happens with positive probability. Separate $m(c + 1)$ of the central agents into $m$ groups of $c + 1$ agents. Within each group, the agents can be made to be connected in a clique with all the other group members if the type distance between the two greatest separated members of that group is smaller than the type distance of any member of that group to the closest type member of any other group. Thus there is a realization of types such that all the $m$ groups are composed of cliques. Now consider the remaining $r$ agents that were not placed in a clique. Since all the other agents outside of these $r$ agents are already at capacity, and given that $r < c - 1$, it is not possible for these $r$ agents can reach capacity only by linking with each other. Therefore there must be an available link to an extreme agent.

Next suppose that $K$ is divisible by $c + 1$. Suppose that $K$ is not divisible by $2c$, and let $n$ be the greatest integer such that $n(2c) < K$. Let $r' = K - n(2c)$, and suppose for now that $r' \neq c + 1$. Take $n(2c)$ of the agents and divide them into $n$ groups of $2c$ agents each. We show that within each group the agents can be made to reach capacity. This can be done by first assuming that the agents within each group are closer to each other in type than they are to any other agent of any other group. Take the $2c$ agents in each group and further divide them into two subgroups of $c$ agents each. If within each subgroup the types of the $c$ agents are closer to each other than to any member of the other subgroup, the agents in each subgroup will form a clique with each other, and they will all be $1$ link under capacity afterwards. Then, suppose that for concreteness the agents in the second subgroup all have types greater than the agents of the first subgroup. To fill their capacities, the agents in the first subgroup will link with the agents in the second subgroup, starting with the greatest in type agent of the first subgroup and the smallest in type agent of the second subgroup, then the second greatest agent of the first subgroup and the second smallest agent of the second subgroup, and so on until all the agents reach their respective capacities.

Suppose that $r' < c + 1$. Then the remaining $r'$ agents, who can only form links with each other, will be unable to form enough links to satisfy their capacities and so there will be an available link. Next suppose that $r' > c + 1$. In this case we can split off $c + 1$ agents from this set of remaining agents and have them form a clique. The agents that were not split off will number less than $c + 1$ and so be unable to fulfill their capacities, resulting in an available link.

Now suppose still that $K$ is divisible by $c + 1$ and not $2c$, but now the remainder $r'$ defined above is equal to $c + 1$. In this case, the above method can be modified to still produce an available link. Take $(n - 1)(2c)$ of the central agents and divide them into $n - 1$ groups of $2c$ agents, and take $2(c + 1)$ of the other central agents and divide them into two groups of $c + 1$ agents. Following the same procedure as above, the agents within each group can be made to reach capacity. Afterwards, there will be $c - 1$ agents that were not placed into a group, and these $c - 1$ agents cannot reach capacity by linking only with themselves and so there must be an available link.

Finally, suppose that $K$ is divisible by both $c + 1$ and $2c$. Let $n = \frac{K}{2c}$. A similar method as above can be adopted. Take $(n - 1)(2c)$ of the central agents and divide them into $n - 1$ groups of $2c$ agents, and take $(c + 1)$ of the other central agents and form a clique with these agents. All the agents that were chosen in this way can be made to hit capacity, and there will be exactly $c - 1$ leftover agents. These $c - 1$ agents cannot hit capacity only by linking amongst themselves, and so there will be an available link.

∎

This result is interesting in that it predicts different behaviors of the final network depending on the precise number of agents as well as the capacity. If the capacity is very low and equal to $1$, there will be no or very few available links, and if the capacity is large and equal to $K - 1$ there will be no available links. For all other capacity values there may be more available links possible. We test this result via simulations. First we assume that there are 7 central agents with prior distributions drawn from a $N(10, .01)$ distribution, and 2 extreme agents, one from a $N(9, .01)$ distribution and one from a $N(11, .01)$ distribution. (We use normal distributions for greater tractability in the simulations, but with low initial variances the two settings should be similar, as with high probability all the central agents will link with each other before linking with an extreme agent.) We assume that the tolerance is very large, equal to 10, such that the extreme agents will still connect with each other even if they cannot connect with a central agent. We vary the capacity of the agents from 1 to 7, and we plot the average number of links for the two extreme agents as a function of the capacity. The results are shown in Figure 9.

Interestingly, the simulation shows that the average number of links exhibits strong non-monotonicity depending on the capacity. The average number of links increases at first, but then decreases when capacity gets larger with a minimum at 6. The capacity of 6 is a special value because it represents the capacity at which the central agents will form a clique with high probability and the extreme agents can only link with each other. However when the capacity is 7, now each central agent will have one available link so the extreme agents can link to each central agent for sure.

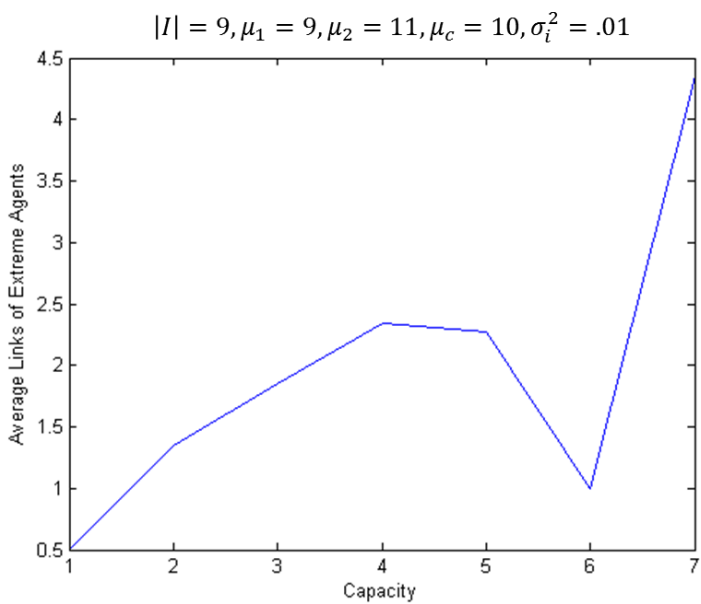

Figure 9: Average Links of Extreme Agents as a Function of Capacity with Seven Central Agents

We repeat the same simulation but we add an extra central agent so that there are now an even number of central agents. This is shown in Figure 10. Fitting in line with our earlier results, the pattern is still non-monotonic and changes greatly with the capacity. It increases somewhat initially and then decreases at the capacity of 7, which is the capacity at which all central agents are likely to be connected in a clique without spots for the extreme agents. It then increases greatly at the capacity of 8 since all the central agents will now have at least 1 open spot. The non-monotonic nature of the graph shows the fragility of these networks. Overall the main takeaway is that extreme agent networks can be very sensitive to the precise specifications, as slight changes in capacities can result in drastically different outcomes in the network structure.[10]

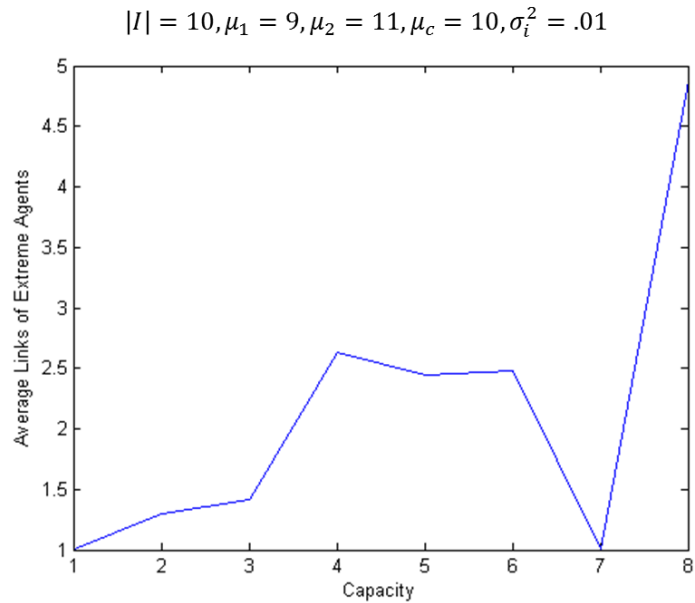

Figure 10: Average Links of Extreme Agents as a Function of Capacity with Eight Central Agents

We repeat the above simulations, but this time with a larger number of agents, to see what types of trends manifest as the set of agents becomes larger. Figure 11 shows the average number of links when there are 50 and 51 central agents respectively. A number of features stand out. First, both graphs are much smoother than the case with very few agents, and display a notable trend which is an initial increase to a peak followed by a decrease. The graphs are also much more similar to each other than is the case with only 7 or 8 agents, as the peaks and valleys of the graphs occur at very similar places. Finally, the overall trend of the graphs indicates that extreme agents are initially helped by the central agents having a higher capacity, but eventually hurt as the capacity of the central agents becomes too large. Then when the capacity no longer binds, which occurs at the values of 50 and 51 respectively, the extreme agents are once again able to form many links.

---

[10] We note that the capacity is changing simultaneously for all agents, so in that sense any change in the capacity level may be expected to have a significant effect. If the network has many central and extreme agents and the capacity was only changed for a few agents, then the impact should be relatively minor.

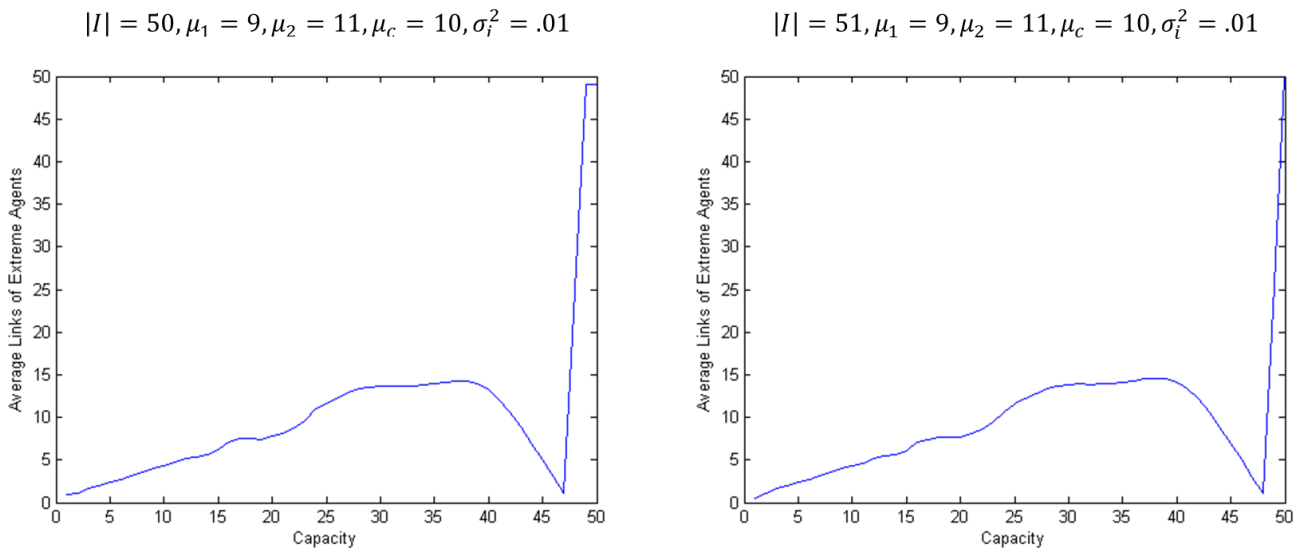

Figure 11: Average Links of Extreme Agents as a Function of Capacity with **50** and **51** Central Agents

*Bimodal Distribution*

We now repeat the analysis above using a bimodal normal distribution in order to test the robustness of our results. The previous simulation analysis will be redone with a bimodal normal distribution, with a **50%** weight on a $N(10,1)$ distribution, and a **50%** weight on a $N(7,2)$ distribution. A nice property of a bimodal normal distribution is that the Bayesian updating remains computationally simple. Since there is a one half probability of coming from the first distribution and a one half probability of coming from the second distribution, the agent beliefs at each time step can be computed twice assuming that each one of the distributions is correct. This updating process utilizes the tractable Bayesian updating formula for normal distributions, and the actual belief at each time step will be a weighted average of these two beliefs. The resulting graphs show similar general features as in the original unimodal normal distribution case, confirming the robustness of the initial results.

Figure 12 shows the Jaccard distance between the final complete and incomplete network, similar to Figure 3. As in that figure, the Jaccard distance is zero or infinity at large or small social distances, and reaches a single peak at an intermediate value. Although the shape is very similar, one difference is that the social tolerance level where the Jaccard distance is the greatest is at a larger social tolerance level than in the unimodal distribution. In addition, the maximum Jaccard distance is also lower.

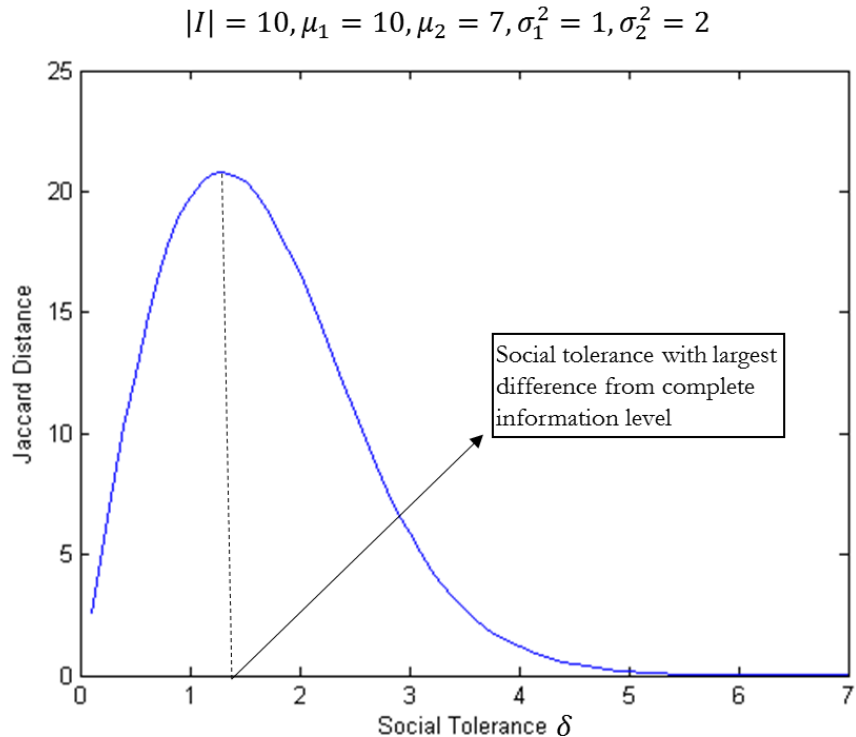

Figure 12: Jaccard Distance between the final complete and incomplete information network with bimodal distribution

Figure 13 shows the average number of links in the final incomplete information network at different social tolerance levels with the bimodal distribution. Once again the figure looks very similar to the previous Figure 5. There is an S shape with the average number of links increasing sharply at first and then plateauing after the tolerance becomes large. The plateau level is very similar to that of Figure 5, but the inflection point is at a higher social tolerance level than before.

Finally we note that analogues of Figures 6 and 7 showing the average total distance and the difference in local clustering between the final complete and incomplete information networks can also be computed for the bimodal distribution. Like the other figures the resulting graphs look very similar, confirming the robustness of these results as well, and we avoid including them for the sake of brevity.

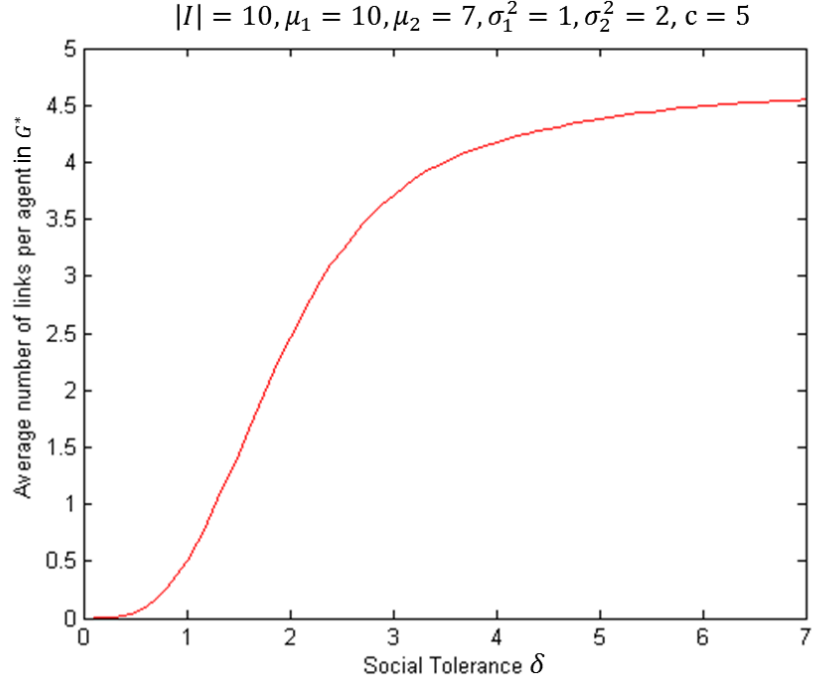

Figure 13: Average Number of Links in $G^*$ Based on Social Tolerance with Bimodal Distribution

*Different Stages of Life*

We now analyze the predictions made by our model when agents and network formation pass through different stages of life. As discussed previously, we iterate the experimentation phase and learning phase multiple times, with each agent's true type subjected to a normal noise term at the end of each stage. In the figure below, we do this for a total of 30 stages, and the variance of the type evolution noise is given by $\frac{1}{2^t}$, where $t$ represents the stage. This assumes that each agent's type changes less as they age. This is consistent with psychological research, which shows that agent personality traits continue to evolve through their entire lives, but that the largest changes tend to happen at younger ages (Roberts and Mroczek 2008). Since the variance of the type evolution decreases exponentially over time, both the complete information and incomplete information networks will eventually converge.

Figure 14 shows the Jaccard distance over these different stages of life at different capacity levels. First note that for each capacity level, the distance increases initially as each agent's type is subjected to the high levels of randomness. With complete information the agents will always know each others' types and so find the appropriate partners to link with even with this randomness, but with incomplete information the randomness will cause many links to be broken. Thus the two networks drift apart from each other at the start due to this randomness. Eventually however, since the randomness decreases in later stages, both networks will converge to a final configuration, with the Jaccard distance decreasing. For higher levels of capacity, greater differences in the links between the complete and incomplete information networks are possible, and so the Jaccard distance increases as the capacity increases.

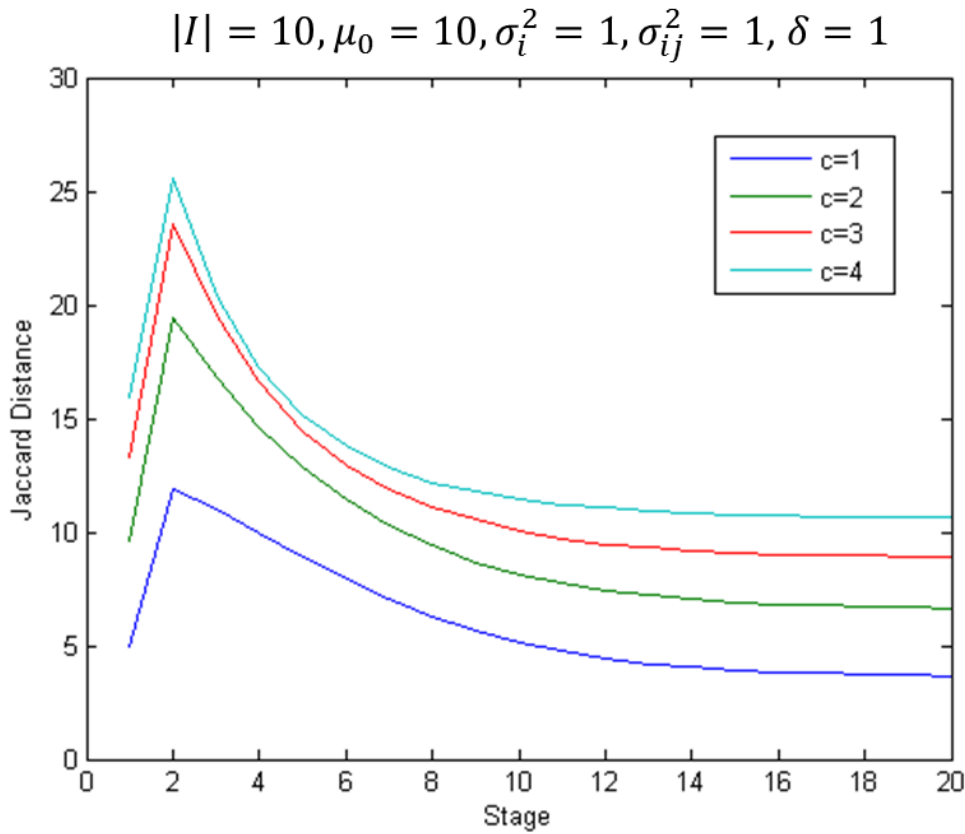

Figure 14: Jaccard distance between complete and incomplete information network for different stages of life and at different capacity levels.

Figure 15 shows the Jaccard distance over different stages of life at different social tolerance levels. For all social tolerance levels, the shape of the curve is similar to the shapes in Figure 14 with a single peak and then a decline. However, the Jaccard distance does not change monotonically with the social tolerance as it does with the capacity. The lowest social tolerance of .1 gives the lowest Jaccard distance throughout, but the highest social tolerance of 1 starts and ends at an intermediate Jaccard distance, although it does feature the largest peak Jaccard distance. This non-monotonicity of the Jaccard distance with respect to social tolerance mirrors the non-monotonicity that is present in Figure 3. At larger social tolerance levels more links are possible so the potential differences in the links is greater, but learning also becomes more accurate which tends to reduce the differences between complete and incomplete information. Even with multiple stages of life, an analogue of Theorem 3 holds so that as the social tolerance becomes very large or very small the Jaccard distance must approach zero at all stages of life.

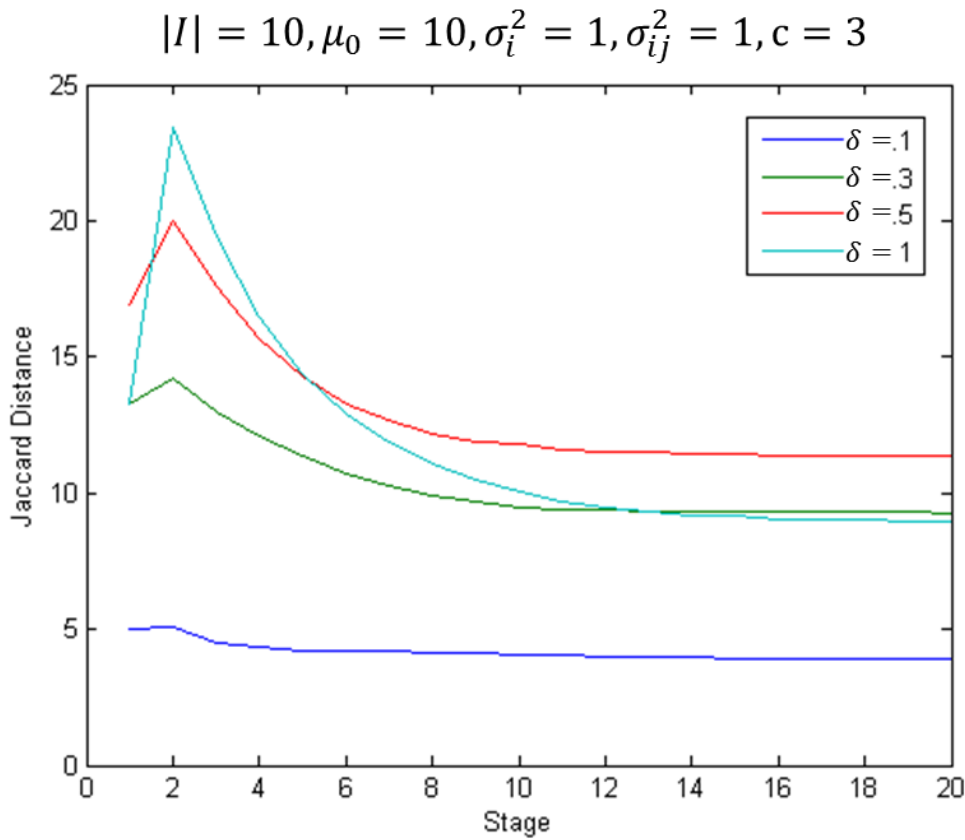

Figure 15: Jaccard distance between complete and incomplete information network for different stages of life and at different social tolerance levels.

## 10. Conclusion

In this paper we formulated a model to capture the impact of homophily on the interaction between learning and network formation. In this model, agents learn about each other through mutual interactions; homophily leads them to maintain connections only with agents that are (learned to be) similar to them. Our results show that incomplete information leads to lower clustering than would arise under complete information. The reason is that due to errors in the learning process, agents may not actually find out which others are most like them, so with some probability agents may choose neighbors who are less similar to themselves than the neighbors they would choose when information is complete. The extent of the difference between the complete and incomplete information network depends heavily on social tolerance; surprisingly, the difference goes to zero as the social tolerance becomes very small or very large. When we incorporate multiple stages of life, we see that the difference between the complete and incomplete information networks is initially very large, but eventually decreases over time.

Many interesting extensions suggest themselves. One obvious extension is to allow the signals that the agents send to their neighbors in each time period to be correlated (rather than independent). Such correlation may lead to greater clustering, because when one neighbor receives an extreme signal and so severs a link, other neighbors will be more likely to receive an extreme signal and also sever the link. Another obvious extension is to allow for indirect learning via friends of friends. If agent A is connected to agent B and agent B is connected to agent C, then agent A might learn about agent C indirectly -- even if they are not directly linked (although presumably indirect learning would be noisier). Indirect learning might also lead to greater levels of clustering, as friends of friends are now more likely to find out about each other and hence connect.

Potentially interesting – and realistic -- extensions of the multiple stages of life model would be to allow for changes in agent characteristics across different stages of life. For instance, agents might become either more or less tolerant as they age, which could have a large impact on the Jaccard distance between complete and incomplete information networks: if agents begin life very tolerant but become less tolerant over time, the Jaccard distance might increase as agents age, rather than decreasing. Changes in tolerance – and in other characteristics – might be exogenous, but endogenous changes – changes that are influenced on the basis of links that are formed – seem especially interesting and important.

Finally, it would be very interesting to use data from real world networks to test the predictions of the model. For instance, we might use data of networks of coauthors in academic journals to see how ties between authors evolve over time. In this setting the type could represent the academic interests of the author and the signals could be previous publications of the authors. Alternatively, we might test the predictions of our model in online social networks (in which type would represent the interests of the agents and the signals would be the messages that agents send to each other) to see how links between friends change as agents learn and interact with each other.

Monte Carlo Simulation Technical Appendix

We perform Monte Carlo simulations of the network learning and formation process to derive properties of the resulting networks. Fixing the set of $I$ agents and the normal prior distributions $N(\mu_i^0, \sigma_i^2)$, for each run of the simulation we take a draw of true agent types from the prior distribution. We then form the initial network $G^0$ based on the tolerances $\delta_i$ and these true types. Using this network $G^0$, we simulate the signals that each agent sends in each period of the experimentation phase. All the simulations in the paper use one period for the experimentation phase, and thus one signal is sent per agent during experimentation. Greater numbers of periods will tend to increase the differences with the complete information network. From the agent signals, we can compute the learned information network $G^L$ that forms using the agent linking rules.

For each $G^L$ we then apply the capacity constraints using the stable network procedure described in the paper to get the final network $G^*$. Thus in each run of the simulation we can arrive at a final network $G^*$.

Using the same types that were drawn initially, we can also derive the resulting complete information network $G^C$ by computing which agents wish to link with each other given the social tolerance thresholds. We then apply the capacity constraints onto the network to get the final complete information network $G_C^*$. From all the runs of the distribution, we can then get estimates of the probability that each final incomplete information network $G^*$ and complete information network $G_C^*$ will form.

From the distributions of $G^*$ and $G_C^*$ obtained through the runs of the simulation, we can analyze different parameters of interest, which are defined below.

Definitions:

**Average number of links of each agent**

From the numerical analysis, we get the probability of the network evolving into each possible final network. Then we take the number of links each agent has in each final network, and we weight it by the probability that final network occurs, to get the average number of links each agent has.

**Jaccard Distance**

For a fixed final complete information and final incomplete information graph, we count the number of differences in the adjacency matrices of the two graphs and sum them up. We take the average of this sum over all runs of the Monte Carlo simulations.

**Average Total distance of links of each agent**

The sum of the distance of all the links each agent has in each final network. We take the average over the agents.

**Local Clustering Coefficient**

The average of the local clustering coefficient of all the nodes of the network. The local clustering coefficient of a node is equal to the proportion of links that exist among the subgraph of that node and its direct neighbors divided by the total number of links that are possible within this subgraph.